\documentclass[conference]{IEEEtran}
\IEEEoverridecommandlockouts
\usepackage{cite}
\usepackage{amsmath,amssymb,amsfonts}
\usepackage{algorithmic}
\usepackage{graphicx}
\usepackage{textcomp}
\usepackage{xcolor}
\usepackage{multirow}
\usepackage{booktabs}
\usepackage{url}
\usepackage{flushend}

\def\BibTeX{{\rm B\kern-.05em{\sc i\kern-.025em b}\kern-.08em
    T\kern-.1667em\lower.7ex\hbox{E}\kern-.125emX}}
\begin{document}

\title{Mamba-Diffusion Model with Learnable Wavelet for Controllable Symbolic Music Generation\\

}

\author{\IEEEauthorblockN{1\textsuperscript{st} Jincheng Zhang}
\IEEEauthorblockA{\textit{Centre for Digital Music} \\
\textit{Queen Mary University of London}\\
London, UK \\
jincheng.zhang@qmul.ac.uk}
\and
\IEEEauthorblockN{2\textsuperscript{nd} Gy\"orgy Fazekas}
\IEEEauthorblockA{\textit{Centre for Digital Music} \\
\textit{Queen Mary University of London}\\
London, UK \\
george.fazekas@qmul.ac.uk}
\and
\IEEEauthorblockN{3\textsuperscript{rd} Charalampos Saitis}
\IEEEauthorblockA{\textit{Centre for Digital Music} \\
\textit{Queen Mary University of London}\\
London, UK \\
c.saitis@qmul.ac.uk}
}

\maketitle

\begin{abstract}

The recent surge in the popularity of diffusion models for image synthesis has attracted new attention to their potential for generation tasks in other domains. However, their applications to symbolic music generation remain largely under-explored because symbolic music is typically represented as sequences of discrete events and standard diffusion models are not well-suited for discrete data. We represent symbolic music as image-like pianorolls, facilitating the use of diffusion models for the generation of symbolic music. Moreover, this study introduces a novel diffusion model that incorporates our proposed Transformer-Mamba block and learnable wavelet transform. Classifier-free guidance is utilised to generate symbolic music with target chords. Our evaluation shows that our method achieves compelling results in terms of music quality and controllability, outperforming the strong baseline in pianoroll generation. Our code is available at https://github.com/jinchengzhanggg/proffusion. 








\end{abstract}


\section{Introduction}



In recent years, diffusion models have been increasingly used for various generation tasks such as image  \cite{b5} and  music  \cite{b4} synthesis. Compared to audio, one advantage of symbolic music is its interpretability as it is represented through fundamental musical elements such as pitch, duration, and velocity. Additionally, symbolic music generated by deep learning models can be easily edited using existing composition tools \cite{b6}. However, symbolic music is often represented as sequences of discrete MIDI tokens, which results in the limited application of diffusion models for their generation as standard diffusion models are not well-suited for handling discrete data \cite{b14}.

Generative models have recently been significantly improved for producing symbolic music \cite{b3}. Despite the recent progress, controllable symbolic music generation remains challenging. There is a growing need of controllability in music generative models. It allows the interaction between diverse users and these models, enabling the generation of music that adheres to the desired musical rules, such as chord progression \cite{b20}. 

Since diffusion models have achieved high-quality and controllable generation of images in computer vision field \cite{b5}, we use this class of generative models to approach controllable symbolic music generation. Our diffusion model uses a U-Net integrated with Transformer \cite{b21}, Mamba \cite{b17}, and wavelet transform as the denoising network. While Transformers rely on self-attention whose computational complexity scales quadratically with the sequence length, Mamba built upon state space model introduces a selection mechanism that allows it to selectively focus on relevant information within input sequences. Wavelet transform is a signal processing method that decomposes a signal into several components at different frequency bands. We transformed MIDI to pianoroll representations, which resemble image formats with time on the horizontal axis and pitch on the vertical axis \cite{b20}, enabling the use of diffusion models for generating symbolic music. Image-like pianorolls are used to train our controllable diffusion model, which learns to progressively denoise starting from Gaussian noise. 

Our main contributions can be summarized as follows:

\begin{itemize}
\item We present a hybrid Transformer-Mamba block to explore the potential of Mamba for improving the generation of symbolic music.
\item To the best of our knowledge, this study is the first attempt that extends learnable wavelet transform to symbolic music generation. 
\item A novel diffusion model is proposed using the hybrid Transformer-Mamba block and learnable discrete  wavelet transform. Our proposed model is capable of providing users with controls to generate symbolic music with target chords.
\end{itemize}

\section{Related Work}

\subsection{Controllable Music Generation}
In the past few years,  many attempts have been made in controllable music generation. Generally speaking, infilling refers to the task of music segment generation controlled by the past and future contexts. Music SketchNet \cite{b9} is capable of
generating the missing measures of incomplete monophonic musical pieces given the surrounding music. It also provides the users with more controls to specify the pitch and rhythm. Guo et al. \cite{b10} presented a transformer-based framework for music infilling. This generative framework is also extensible for additional control signals such as tonal tension per bar. TeleMelody \cite{b13} based on Transformers allows
lyric-to-melody generation. 
Copet et al. \cite{b1} introduced a single-stage transformer Language Model (LM) that
can generate music conditioned on text and melody. FIGARO \cite{b29} based on Transformer is able to generate music that aligns well with both input expert descriptions and learned features. Our work explores the use of diffusion models for symbolic music generation controlled by chords.
\subsection{Diffusion Models for Symbolic Music Generation}

The generation of symbolic music has not been thoroughly empowered by diffusion models. Recent efforts include Mittal et al. \cite{b14}, who trained a diffusion model on latent representations of symbolic music learned by MusicVAE \cite{b15} for the generation of melody. Min et al. \cite{b16} used an image-like pianoroll representation of music which allows directly employing their diffusion models for symbolic music generation without the need of encoding music to latent representations. Huang et al. \cite{b20} trained
a diffusion model with a transformer backbone on a pre-trained pianoroll VAE's latent space. Zhang et al. \cite{b12} proposed a discrete diffusion model combined with a VQVAE to generate pianorolls with target composer styles. Our diffusion model for pianoroll generation conditioned by chords distinguishes itself by implementing a novel denoising U-Net with wavelet transform and Mamba.


\section{Method}

\subsection{Preliminaries}
\paragraph{Diffusion Models}

\begin{figure*}
\centering
\includegraphics[width=0.86\textwidth]{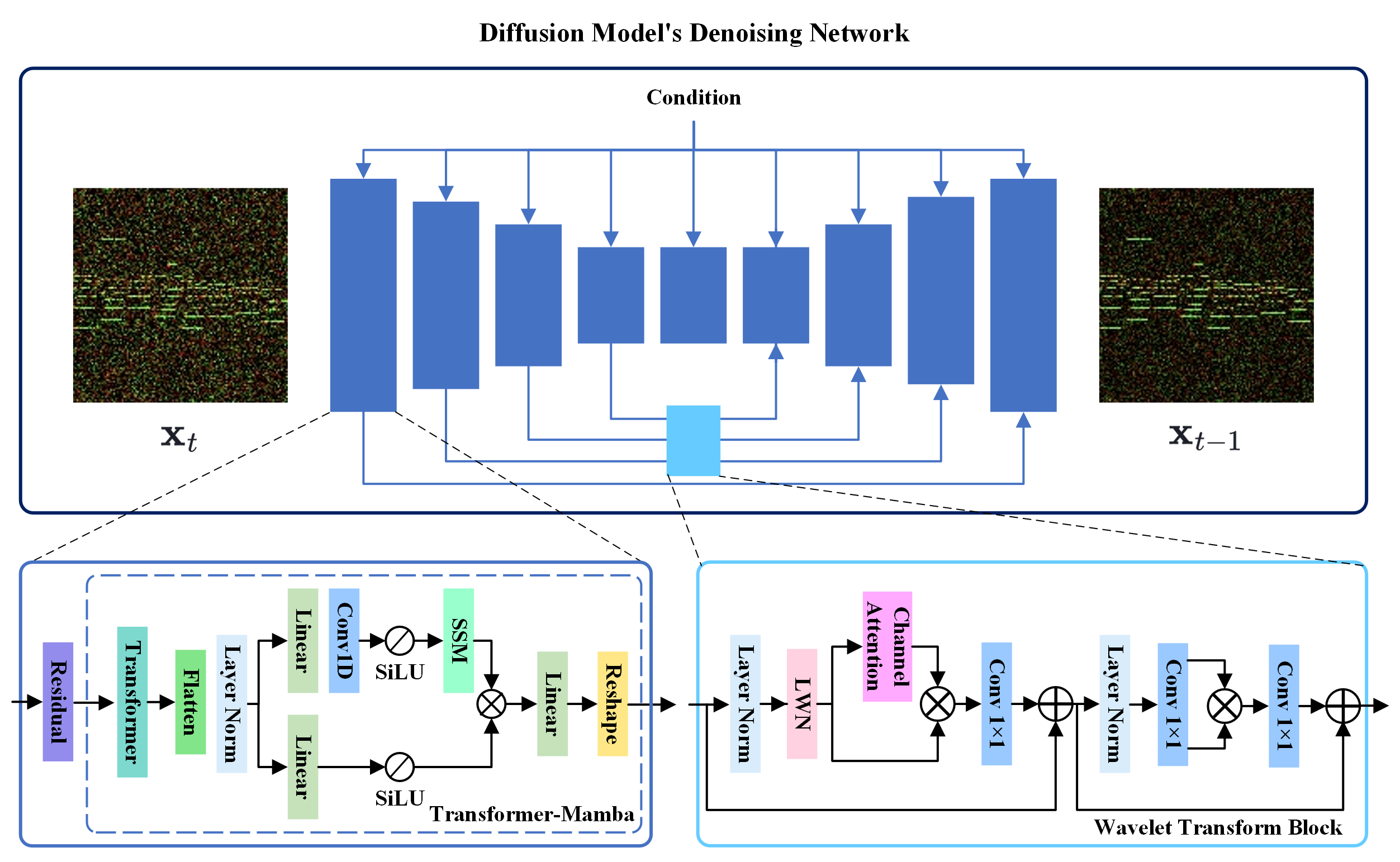}
\caption{\label{fig:vqdiff} Architecture of our proposed Mamba-Diffusion model with Wavelet for controllable music generation. Its denoising network is a U-Net combining learnable wavelet transform and our Transformer-Mamba blocks. Chords are
encoded using a pre-trained VAE and fed into the U-Net via cross-attention
layers.}
\end{figure*}

\begin{figure}[t]
\centering
\includegraphics[width=0.46\textwidth]{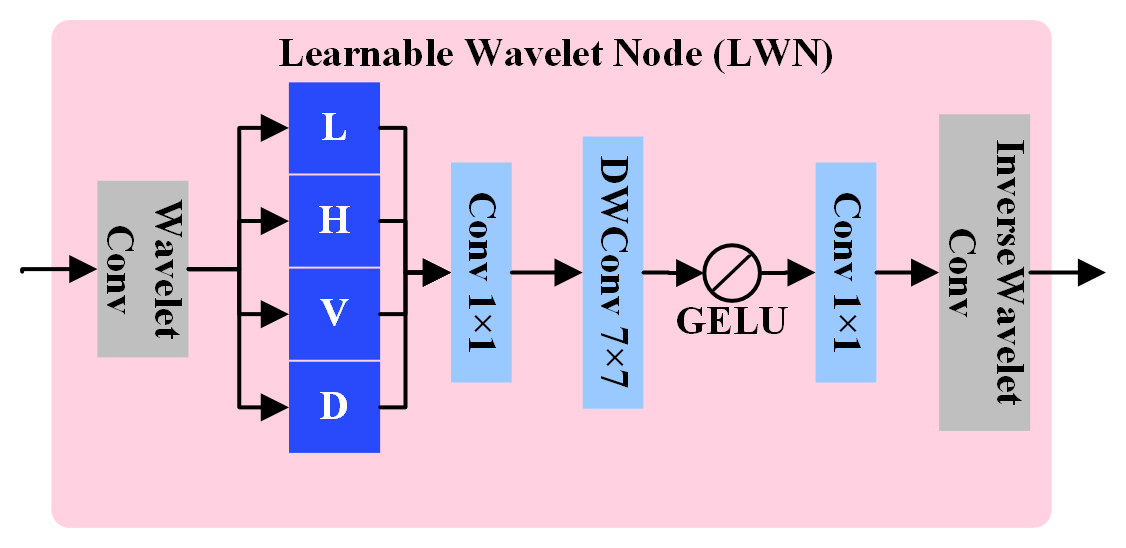}
\caption{\label{fig:boxplot1}
 Illustration of the Learnable Wavelet Node (LWN).}
\end{figure}

Diffusion models have demonstrated impressive capabilities in generating high-quality samples. These generative models are comprised of a forward (diffusion) process and a reverse (denoising) process. The forward process corrupts the original data $\mathbf{x}_{0}$ towards a prior distribution $p\left(\mathbf{x}_{T}\right)$ that is independent of $\mathbf{x}_{0}$ by iteratively adding noise to $\mathbf{x}_{0}$ for $\boldsymbol{T}$ steps,  yielding a sequence of increasingly noisy latent variables $\mathbf{x}_{1}$, \ldots, $\mathbf{x}_{T}$:
\begin{equation}
q\left(\mathbf{x}_{t} \mid \mathbf{x}_{t-1}\right)=\mathcal{N}\left(\mathbf{x}_{t} ; \sqrt{1-\beta_{t}} \mathbf{x}_{t-1}, \beta_{t} \mathbf{I}\right)
\end{equation}
\begin{equation}
q\left(\mathbf{x}_{1: T} \mid \mathbf{x}_{0}\right)=\prod_{t=1}^{T} q\left(\mathbf{x}_{t} \mid \mathbf{x}_{t-1}\right)
\end{equation}
where $\beta_{1}, \ldots, \beta_{T}$ are a series of variance scheduling parameters controlling the amount of noise added in every step. $\mathbf{x}_{0}$ denotes the clean pianorolls in our case. Diffusion models aim to learn a reverse process $p_{\theta}\left(\mathbf{x}_{t-1} \mid \mathbf{x}_{t}\right)$ that gradually denoises to approximate the data distribution.
\begin{equation}
p_{\theta}\left(\mathbf{x}_{t-1} \mid \mathbf{x}_{t}\right)=\mathcal{N}\left(\mathbf{x}_{t-1} ; \mu_{\theta}\left(\mathbf{x}_{t}, t\right), \mathbf{\Sigma}_{\theta}\left(\mathbf{x}_{t}, t\right)\right)
\end{equation}
\begin{equation}
p_{\theta}\left(\mathbf{x}_{0: T}\right)=p\left(\mathbf{x}_{T}\right) \prod_{t=1}^{T} p_{\theta}\left(\mathbf{x}_{t-1} \mid \mathbf{x}_{t}\right)
\end{equation}
The model is trained by minimising the negative log-likelihood of $p_\theta\left(\mathbf{x}_0\right)$ via the variational lower bound (VLB):
\begin{equation} \label{eq:6}
\begin{split}
&\mathbb{E}_{q}[{D_{\mathrm{KL}}\left(q\left(\mathbf{x}_{T} \mid \mathbf{x}_{0}\right) \| p\left(\mathbf{x}_{T}\right)\right)}
{-\log p_{\theta}\left(\mathbf{x}_{0} \mid \mathbf{x}_{1}\right)}\\
+&\sum_{t>1} {D_{\mathrm{KL}}\left(q\left(\mathbf{x}_{t-1} \mid \mathbf{x}_{t}, \mathbf{x}_{0}\right) \| p_{\theta}\left(\mathbf{x}_{t-1} \mid \mathbf{x}_{t}\right)\right)}]
\end{split}
\end{equation}
The reverse process often uses a neural network such as a U-Net \cite{b7} to learn the conditioned probability distribution $p_{\theta}\left(\mathbf{x}_{t-1} \mid \mathbf{x}_{t}\right)$. U-Net features down-sampling and up-sampling blocks, as well as skip connections that facilitate information transfer between corresponding layers of the down-sampling and up-sampling blocks.


\paragraph{State Space Models}

State Space Sequence Models (SSMs) based on  continuous systems map a one-dimensional input stimulation $x(t)$ to response $y(t)$ through a hidden state $h(t)$, which can be formulated as linear ordinary differential equations (ODEs):
\begin{equation}
\begin{aligned}
h^{\prime}(t) & =\mathbf{A} h(t)+\mathbf{B} x(t) \\
y(t) & =\mathbf{C} h(t)
\end{aligned}
\end{equation}
where A is the state transition matrix, while B
and C denote the projection parameters. Structured state space sequence models (S4) discretize
the aforementioned continuous system typically using the zero-order hold (ZOH). The discretized system can
then be defined as follows:
\begin{equation}
\begin{aligned}
\overline{\mathbf{A}} & =\exp (\boldsymbol{\Delta} \mathbf{A}) \\
\overline{\mathbf{B}} & =(\boldsymbol{\Delta} \mathbf{A})^{-1}(\exp (\boldsymbol{\Delta} \mathbf{A})-\mathbf{I}) \cdot \boldsymbol{\Delta} \mathbf{B} \\
h_t & =\overline{\mathbf{A}} h_{t-1}+\overline{\mathbf{B}} x_t \\
y_t & =\mathbf{C} h_t
\end{aligned}
\end{equation}
where $\boldsymbol{\Delta}$ denotes the timescale parameter. $\overline{\mathbf{A}}$ and $\overline{\mathbf{B}}$ are the discrete counterparts of parameters A and B. Given the input sequence $\mathbf{x}$ with the length $\boldsymbol{L}$, global convolution can be used to compute the output with a structured
convolutional kernel $\overline{\mathbf{K}}$:
\begin{equation}
\begin{aligned}
\overline{\mathbf{K}} & =\left(\mathbf{C} \overline{\mathbf{B}}, \mathbf{C} \overline{\mathbf{A B}}, \ldots, \mathbf{C} \overline{\mathbf{A}}^{\boldsymbol{L}-1} \overline{\mathbf{B}}\right) \\
\mathbf{y} & =\mathbf{x} * \overline{\mathbf{K}}
\end{aligned}
\end{equation}
Mamba \cite{b17} further extends S4 by employing a input-dependent selection mechanism which
allows the dependence of the model’s parameters B, C and $\boldsymbol{\Delta}$ on the inputs and selectively filters out irrelevant information. Additionally, Mamba can perform parallel computations with a complexity that scales linearly with the sequence length by using an efficient scan algorithm.

\paragraph{Classifier-free Guidance}
Classifier guidance \cite{b5} enables diffusion models to generate samples from $p\left(\mathbf{x} \mid \mathbf{y}\right)$ given a condition $\mathbf{y}$, by training a classifier $p_{t}\left(\mathbf{y} \mid \mathbf{x}_{t}\right)$ on the pairs of noisy samples and their corresponding conditions. During sampling, the gradient of the log likelihood of this classifier is combined with the score estimate of the diffusion model to guide the generation process:
\begin{equation}
\nabla_{\mathbf{x}_{t}} \log p\left(\mathbf{x}_{t}\right)+\omega \nabla_{\mathbf{x}_{t}} \log p_{t}\left(\mathbf{y} \mid \mathbf{x}_{t}\right)
\end{equation}
where $\omega$ is the guidance scale that controls the strength of the classifier guidance. 

However, classifier guidance necessitates training an extra classifier on the noisy data. We use classifier-free guidance \cite{b22} in this work, which avoids training a classifier by jointly training a conditional and an unconditional diffusion model. Their score estimates are mixed during sampling: 
\begin{equation}
(1+\omega) \nabla_{\mathbf{x}_{t}} \log p_{t}\left(\mathbf{x}_{t} \mid \mathbf{y}\right)-\omega \nabla_{\mathbf{x}_{t}} \log p_{t}\left(\mathbf{x}_{t}\right)
\end{equation}

\subsection{Mamba-Diffusion Model with Wavelet Transform}

U-Net has been commonly used as the denoising network of diffusion models.
We propose a novel diffusion model using a learnable discrete wavelet transform and Mamba-enhanced U-Net as its backbone for symbolic music generation, as shown in Fig. 1. Both the U-Net's encoder and decoder begin with Residual blocks \cite{b28}, which are subsequently followed by our proposed Transformer-Mamba. The combination of Transformers and Mamba aims to leverages the strengths of both architectures. The Residual block contains Group Normalization, SiLU activation function \cite{b23} and convolutional layers. The Transformer-Mamba block's input features with a shape of $(B, C, H, W)$ are first processed by a Transformer while the shape remains unchanged, where $B$, $C$, $H$, $W$ denote the batch size, channels, height and width, respectively. Before the features enter the Mamba, the output of the Transformer is flattened and transposed to $(B, L, C)$ and then normalized, where the feature length $L = H \times W$. The Mamba consists of two parallel branches. By using linear layers, the feature shapes are expanded to $(B, 2L, C)$ in both branches. While one branch only uses the SiLU activation function after its linear layer, the other branch includes a 1D convolutional layer, the SiLU activation function, and the SSM layer following the linear layer. Subsequently, the features produced by the
two branches are merged with the Hadamard product. Eventually, these merged features are reshaped and transposed back to the original shape $(B, C, H, W)$.

The interest in the integration of wavelet transform with deep neural networks has been growing recently\cite{b2}. The skip connections of U-Net are responsible for passing the features extracted by the encoder block to the decoder stage to combine low-level details with high-level features. We propose using a learnable discrete wavelet transform (DWT) to improve the skip connections of our diffusion model's denoising U-Net, as the wavelet transform can not only promote denoising, but also efficiently capture high-frequency components of abrupt signals (such as note onsets and offsets in pianorolls) and fine details \cite{b25}. In the case of 1D-DWT, given the wavelet function $\psi_{j, k}(t)=2^{\frac{j}{2}} \psi\left(2^{j} t-k\right)$ and scale function $\phi_{j, k}(t)=2^{\frac{j}{2}} \phi\left(2^{j} t-k\right)$, the input discrete signal $t$ at $j_{0}$ can be decomposed into a rich wavelet domain signal:
\begin{equation}
f(t)=\sum_{j>j_{0}} \sum_{k} d_{j, k} \psi_{j, k}(t)+\sum_{k} c_{j_{0}, k} \phi_{j_{0}, k}(t)
\end{equation}
where $j$ is the scaling factor, $k$ denotes the time factor, $d_{j, k}=\left\langle f(t), \psi_{j, k}(t)\right\rangle$ and $c_{j_{0}, k}=$ $\left\langle f(t), \phi_{j_{0}, k}(t)\right\rangle$ represent the detail coefficients (i.e., high-frequency components) and the approximation coefficients (i.e., low-frequency components), respectively.

To apply wavelet transform within deep neural networks, the analysis vectors $\vec{a}_{0}[k]$ and $\vec{a}_{1}[k]$ are introduced to represent the high and low frequency filters, respectively.
\begin{align*}
\vec{a}_{0}[k] & =\left\langle\frac{1}{\sqrt{2}} \phi\left(\frac{t}{2}\right), \phi(t-k)\right\rangle \\ 
\vec{a}_{1}[k] & =\left\langle\frac{1}{\sqrt{2}} \psi\left(\frac{t}{2}\right), \phi(t-k)\right\rangle \tag{12}
\end{align*}
Decomposing the original signal can be considered as recursive convolution of the signal with specific filters using a step size of 2:
\begin{align*}
c_{j+1, p} & =\sum_{k} \vec{a}_{0}[k-2 p] c_{j, k} \\
d_{j+1, p} & =\sum_{k} \vec{a}_{1}[k-2 p] c_{j, k} \tag{13}
\end{align*}

The inverse wavelet transform can be similarly implemented by transposed convolution with synthetic vectors $\vec{s}_{0}$ and $\vec{s}_{1}$. The construction of the 2D-wavelet convolution kernel $\mathcal{K}_{w}$ is formulated as follows:
\begin{align*}
& \mathcal{F}_{l l}=\overrightarrow{a_{0}} \times{\overrightarrow{a_{0}}}^{T}, \mathcal{F}_{h l}=\overrightarrow{a_{1}} \times{\overrightarrow{a_{0}}}^{T} \\
& \mathcal{F}_{l h}=\overrightarrow{a_{0}} \times{\overrightarrow{a_{1}}}^{T}, \mathcal{F}_{h h}=\overrightarrow{a_{1}} \times{\overrightarrow{a_{1}}}^{T}  \tag{14}\\
& \mathcal{K}_{w}=\operatorname{cat}\left(\mathcal{F}_{l l}, \mathcal{F}_{l h}, \mathcal{F}_{h l}, \mathcal{F}_{h h}\right)
\end{align*}
where $\overrightarrow{a_{0}}$ and $\overrightarrow{a_{1}}$ are set as learnable filters; $\mathcal{F}_{l l}, \mathcal{F}_{h l}, \mathcal{F}_{l h}$, and $\mathcal{F}_{h h}$ denote low-frequency, vertical high-frequency, horizontal high-frequency, and diagonal high-frequency convolution operators, respectively. These operators are obtained from the vector outer products.

Then learnable wavelet transform can be implemented as the 2D-wavelet convolution. Given input feature maps $\mathcal{X}_{i n} \in(C, H, W)$, the wavelet convolution produces the projection in the wavelet domain $\mathcal{X}_{\text {out }} \in\left(4 C, \frac{H}{2}, \frac{W}{2}\right)$ which comprises low-frequency, vertical high-frequency, horizontal high-frequency, and diagonal high-frequency components. Subsequently, depth-wise convolution with an expansion factor of $r$ is utilized for wavelet domain feature extraction and transformation. A $1 \times 1$ convolution is then applied for channel expansion and scaling, before a learnable wavelet inverse transform is finally employed to project the wavelet domain features back to the spatial domain. These operations constitute the Learnable Wavelet Node (LWN). Incorporating LWN as a base module, the Wavelet Transform Block (WTB) is implemented following \cite{b27}.

To avoid the wavelet convolution degrades into general convolution, an additional self-supervised wavelet loss is used in this study to ensure the correct learning direction of the wavelet kernel. Therefore,
the total loss function includes diffusion and wavelet losses:
\begin{align*}
L_{\text {wavelet }}\left(\theta_{i}\right) & =\left(\sum_{k}^{N-1}\left(\left\langle a_{0}, s_{0}\right\rangle_{k}+\left\langle a_{1}, s_{1}\right\rangle_{k}\right)-\hat{V}_{\left\lfloor\frac{N}{2}\right\rfloor}\right)^{2} \\
\tag{15}
\mathcal{L}_{\text {total}} & =\mathcal{L}_{\text {wavelet }}\left(\theta_{i}\right)+\mathcal{L}_{\text {diffusion}}
\end{align*}
where $\hat{V}$ denotes a vector with a center position value of two, $k$ represents the filter's position when it performs the convolution, and $\theta_{i}$ refers to the constructed filter.

To achieve chord-controllable music generation, beat-wise chords are first extracted as conditions, using the rule-based approaches \cite{b24}. Each chord is a 36-D encoding comprising three 12-D parts, including a one-hot root vector, a one-hot bass vector and a vector for the multi-hot chroma. Subsequently, a pre-trained chord VAE \cite{b24} is applied to encode each 8-bar chord sequence into a 512-D latent representation which is then fed into the U-Net of our diffusion model via a cross-attention \cite{b21} layer. We use classifier-free guidance \cite{b22} to guide the model towards the desired chords. We denote our \textbf{p}iano\textbf{ro}ll di\textbf{ffusion} model with wavelet transform and mamba as \textbf{Proffusion}-WM.


\section{Experiments}
\subsection{Dataset and Representation}
We trained our diffusion model on the POP909 dataset \cite{b18}, which contains approximately 1000 MIDI files of pop songs. These symbolic music pieces are segmented into 8-bar musical excerpts with a 1-bar hop size, yielding a total of 64K samples. The dataset is randomly split at a song level into 90\% and 10\% for training and validation sets, respectively. 1/4 beat is used as the time step to represent the 8-bar (32-beat) long symbolic music segments as image-like pianoroll, leading to 128 time steps per sample. Besides, we use a pitch range spanning from 0 to 127. Therefore, each pianoroll sample has a shape of $(2, 128, 128)$.

\subsection{Experiment Setup}
\begin{figure}[b]
\centering
\includegraphics[width=0.45\textwidth]{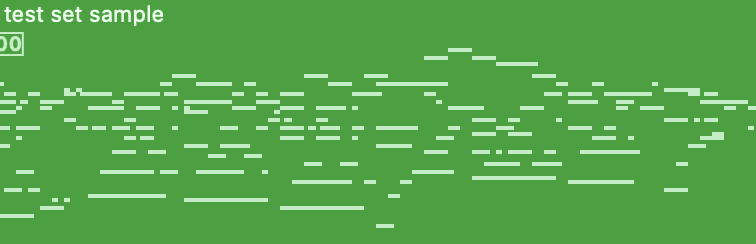}
\caption{\label{fig:testset samples}Pianoroll sample from the test set, with pitch on the vertical
axis and time on the horizontal axis.}
\end{figure}
\begin{figure*}
\centering
\includegraphics[width=0.85\textwidth]{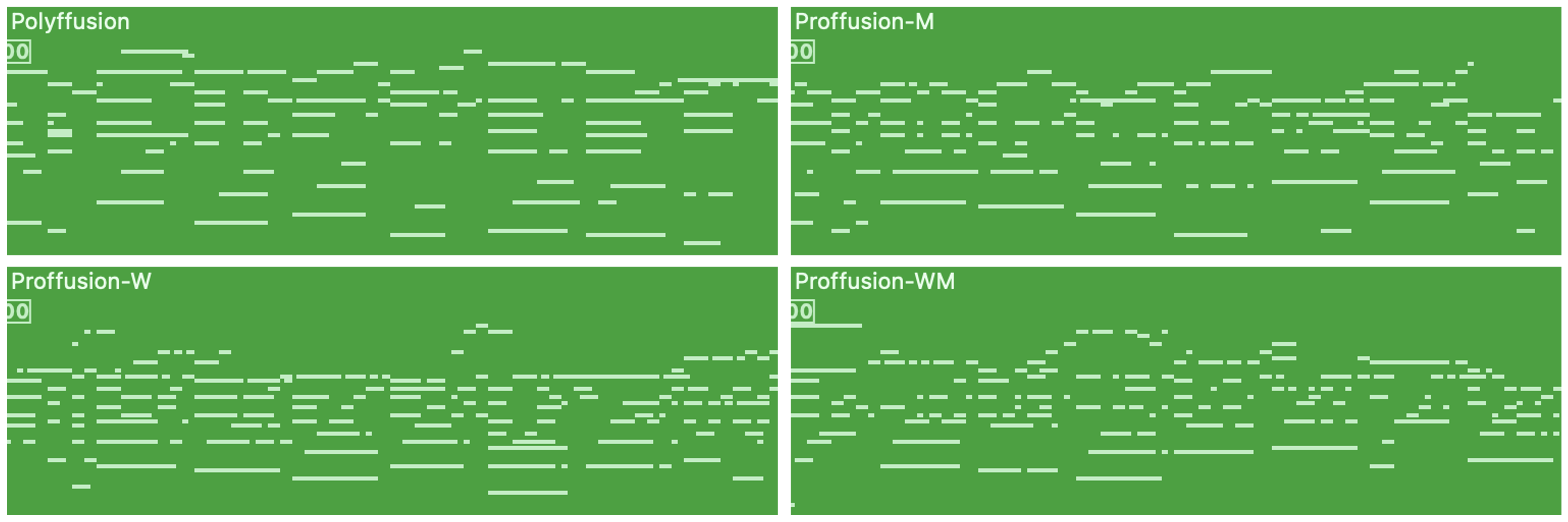}
\caption{\label{fig:music samples}Pianorolls generated by our proposed diffusion model and the other models. }
\end{figure*}

The proposed diffusion model is optimized using Adam Optimizer with a constant learning rate 5e-5. We set the diffusion steps to 1000. Classifier-free guidance with a probability of 0.2 is applied, which randomly drops out the conditioning during training. A guidance scale of 5 is used to generate 2-channel pianoroll samples. Then the generated 8-bar pianorolls are converted back to MIDI files. All models in this work are trained on an RTX 4090 GPU with a batch size of 16.

\subsection{Objective Evaluation}
Quantitative evaluation of music quality remains an open problem. We used the Overlapping Area (OA) metric \cite{b19} to quantify the similarity between the distributions of the generated samples and the ground truth segments. We calculated the OA for musical features including pitch range, duration, inter-onset-interval (IOI) and average OA across the three features. Moreover, F1 score is employed to measure whether the generated samples follow the target chords. The higher values reflect the better results for both the two objective metrics.


\subsection{Listening Test}

We also carried out a listening test to subjectively compare our model with the baselines, complementing the objective evaluation. Eight out of 30 participants in this test are musicians. Each model generated three symbolic music pieces, resulting in a total of 12 samples. The generated symbolic music was then rendered into audio for evaluation. These samples were presented in a random order and their sources were unknown to the participants. We asked participants to listen to the music samples and rate them on a 100-point scale in terms of the five subjective metrics, including Humanness, Harmony, Rhythm, Richness, and Overall Preference. Humanness measures how closely the generated music resembles human compositions; Richness assesses the diversity of the music; Rhythm examines the arrangement and pattern of sounds and silences within the music; Harmony refers to whether multiple notes played simultaneously produce a harmonious sound.

\begin{table*}
\caption{Overlapping Area (OA) similarity between the
distributions of the ground truth segments and the samples generated by different models. For each musical feature, the highest OA values are bolded.}
\centering
\renewcommand{\arraystretch}{1.5}
\resizebox{0.70\textwidth}{!}{
\begin{tabular}{cccccc}
\specialrule{0.8pt}{0pt}{0pt}
 
\multirow{2}{*}{Model} & \multicolumn{4}{c}{OA} & Accuracy             \\ \cline{2-6}
& Pitch Range & IOI & Note Duration & Avg & Chord F1 \\ \hline
 
Polyffusion   & 0.919                & 0.877                                  & 0.790          & 0.862             & 0.494             \\ 
 
Proffusion-M    & 0.948                & 0.940                                  & 0.801           & 0.896           & 0.516             \\ 
 
Proffusion-W  & 0.920                & 0.947                                  & \textbf{0.919}           & 0.929           & 0.517             \\

Proffusion-WM & \textbf{0.955}       & \textbf{0.952}                         & 0.909      & \textbf{0.939}       & \textbf{0.541}    \\
\specialrule{0.8pt}{0pt}{0pt}
\end{tabular} 
}
\label{tab:model_comparison}
\end{table*}

\section{Results and Discussion}

\begin{table*}
\caption{Comparisons of different models for symbolic music generation in terms of subjective metrics. The mean is supplemented with standard error of the mean.}
\centering
\renewcommand{\arraystretch}{1.5}
\resizebox{0.90\textwidth}{!}{
\begin{tabular}{cccccc}
\specialrule{0.8pt}{0pt}{0pt}
Model & Humanness & Richness & Rhythm & Harmony & Overall Preference \\ \hline
Polyffusion & 55.91 $\pm$ 1.31 & 59.09 $\pm$ 1.47 & 54.43 $\pm$ 1.55 & 56.08 $\pm$ 1.58 & 55.43 $\pm$ 1.42   \\
Proffusion-M & 69.59 $\pm$ 1.14 & 71.20 $\pm$ 1.21 & 66.88 $\pm$ 1.35 & 68.96 $\pm$ 1.36 & 70.80 $\pm$ 1.08   \\
Proffusion-W & 64.51 $\pm$ 1.31 & 65.68 $\pm$ 1.41 & 69.33 $\pm$ 1.62 & 71.31 $\pm$ 1.28 & 66.11 $\pm$ 1.31   \\  
Proffusion-WM & \textbf{78.76} $\pm$ \textbf{1.06} & \textbf{73.88} $\pm$ \textbf{1.03} & \textbf{80.43} $\pm$ \textbf{1.14} & \textbf{79.49} $\pm$ \textbf{1.12} & \textbf{80.16} $\pm$ \textbf{1.00}   \\ 
\specialrule{0.8pt}{0pt}{0pt}
\end{tabular}%
}
\label{tab:my-table}
\end{table*}

We compared our diffusion model with Polyffusion \cite{b16} which is also a diffusion model-based method for pianoroll generation. Furthermore, the learnable wavelet transform's effectiveness is evaluated by removing the wavelet transform block in the proposed model. This model without wavelet transform is denoted as Proffusion-M. We also implemented a Proffusion-W model without Mamba for comparison. All the four music generative models were trained on the POP909 dataset. We extracted the chords from all the samples in the POP909 test set, yielding 879 chords which was used as conditions to generate 879 music 8-bar samples for each model. Fig. \ref{fig:testset samples} and Fig. \ref{fig:music samples} show a test set sample and a set of the generated pianorolls, respectively. For demonstrations, please visit our project website: \url{https://proffusion.github.io}.

To ensure an accurate evaluation of OA similarity, it is typically required that the two datasets being compared have an equal number of music samples. We computed the OA for each aforementioned musical feature between the generated music and 879 ground truth segments. The detailed OA results are presented in Table \ref{tab:model_comparison}. Our method Proffusion-WM surpasses the other three models in terms of average OA across the three features. For most individual attributes, it also achieves the best OA, except that its OA for Note Duration is slightly suboptimal compared to Proffusion-W. This is potentially because the addition of mamba to Proffusion-W introduces a trade-off between global and local attributes. The mamba added to the U-Net's encoder and decoder might lead to Proffusion-WM allocates more capacity to the global attributes such as Pitch Range, which could result in less focus on the local attributes and thus a slight degradation of OA for Note Duration. Similar suboptimal results of OA were also observed in the literature \cite{b20}. Overall, the OA results indicate our diffusion model effectively learned the distribution of the musical features from the training data, significantly outperforming the Polyffusion baseline.


The adherence of the generated music to the chord conditions is measured by calculating the Chord F1 scores between the chords of the generated music samples and those fed into the generative models. The Chord F1 results for the four different models are also shown in Table 1. Our proposed Proffusion-WM method obtains the highest chord control accuracy of 0.541, suggesting the chords of the generated music pieces align well with those extract from the ground truth.

We assessed the effectiveness of the learnable wavelet transform by comparing the evaluation results of our Proffusion-WM with those of Proffusion-M. Proffusion-WM achieves higher OA and F1 scores than Proffusion-M. Intuitively, this is likely attributed to the superior capabilities of wavelet transform in image deblurring \cite{b25}, Dehazing \cite{b26} and denoising tasks. Although we use wavelet transform in our model for pianoroll generation task, the nature of the diffusion model's reverse process is progressively denoising randomly sampled Gaussian noise to generate clear pianorolls.


\begin{figure}[!b]
\centering
\includegraphics[width=0.45\textwidth]{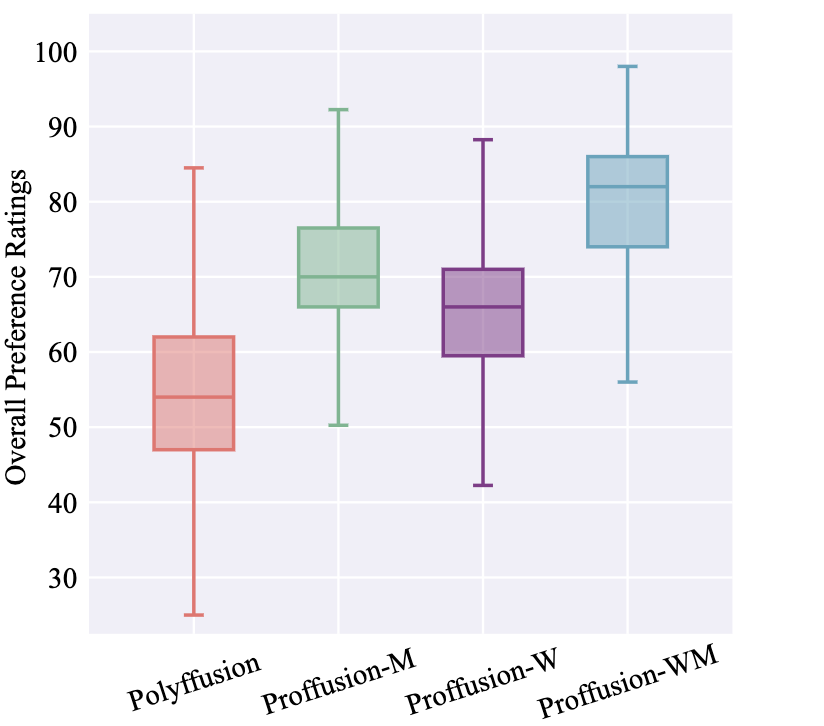}
\caption{\label{fig:boxplot}
Boxplots of Overall Preference ratings for the four different models.}
\end{figure}





Table 2 presents the results of the subjective listening test. For all the metrics, our Proffusion-WM relatively outperforms the Polyffusion, Proffusion-M and Proffusion-W while Polyffusion obtained the worst results. The scores of Rhythm achieved by Proffusion-WM, Proffusion-W, Proffusion-M and Polyffusion decrease sequentially, showing a similar trend with the objective OA results of IOI. This is a good agreement as IOI is strongly related to the human perception of Rhythm. Similarly, Harmony is closely associated with the objective Chord metric. Proffusion-M shows Harmony comparable with Proffusion-W, outperforming Polyffusion but falling short of Proffusion-WM, which is consistent with the Chord F1 results. The rating distribution of the Overall Preference metric is visualized in Fig. \ref{fig:boxplot}. It indicates that our Proffusion-WM model has a higher median rating than the other three models.



\section{Conclusion and Future Work}

In this study, we present a novel mamba-diffusion model enhanced by learnable discrete wavelet transform for the generation of symbolic music. Specifically, our proposed hybrid Transformer-Mamba block is utilized in the encoder and decoder of the diffusion model's denoising U-Net. In addition, the features extracted by the encoder block pass through the wavelet transform block before being fed into the corresponding decoder stage. The comprehensive evaluation results demonstrate that our model outperforms existing state-of-the-art methods for pianoroll generation in terms of general quality and controllability. In the future, we plan to explore different approaches for integrating wavelet transform with denoising U-Net in diffusion models. For instance, it is worth investigating the use of discrete wavelet transform in the U-Net's encoder and inverse wavelet transform in the decoder.

\vspace{12pt}


\clearpage











\begin{thebibliography}{00}


\bibitem{b5} P. Dhariwal and A. Nichol, ‘Diffusion models beat GANs on image synthesis’, in \textit{Advances in Neural Information Processing Systems}, 2021, pp. 8780–8794.

\bibitem{b4} F. Schneider, O. Kamal, Z. Jin, and B. Schölkopf, ‘Moûsai: Efficient text-to-music diffusion models’, in \textit{Proceedings of the 62nd Annual Meeting of the Association for Computational Linguistics}, 2024, pp. 8050–8068.


\bibitem{b6} [1] L. Atassi, ‘Generating symbolic music using diffusion models’, 2023, \textit{arXiv preprint}: arXiv:2303.08385. 


\bibitem{b14} G. Mittal, J. Engel, C. Hawthorne, and I. Simon, ‘Symbolic music generation with diffusion models’, in \textit{Proceedings of the 22nd International Society for Music Information Retrieval Conference}, 2021.

\bibitem{b3} Z. Wang, L. Min, and G. Xia, ‘Whole-song hierarchical generation of symbolic music using cascaded diffusion models’, in \textit{International Conference on Learning Representations}, 2024.

\bibitem{b20} Y. Huang et al., ‘Symbolic music generation with non-differentiable rule guided diffusion’, in \textit{International Conference on Machine Learning}, 2024.

\bibitem{b21} A. Vaswani et al., ‘Attention is all you need’, in \textit{Advances in Neural Information Processing Systems}, 2017. 

\bibitem{b17} A. Gu and T. Dao, ‘Mamba: Linear-time sequence modeling with selective state spaces’, in \textit{COLM}, 2024.

\bibitem{b9} K. Chen, C. Wang, T. Berg-Kirkpatrick, and S. Dubnov, ‘Music SketchNet: Controllable music generation via factorized representations of pitch and rhythm’, in \textit{Proceedings of the 21nd International Society for Music Information Retrieval Conference}, 2020.

\bibitem{b10} R. Guo, I. Simpson, C. Kiefer, T. Magnusson, and D. Herremans, ‘MusIAC: An extensible generative framework for music infilling applications with multi-level control’, in \textit{Artificial Intelligence in Music, Sound, Art and Design}, vol. 13221, 2022, pp. 341–356. 

\bibitem{b13} Z. Ju et al., ‘TeleMelody: Lyric-to-melody generation with a template-based two-stage method’, in \textit{Proceedings of the 2022 Conference on Empirical Methods in Natural Language Processing}, 2022, pp. 5426–5437.

\bibitem{b1} J. Copet et al., ‘Simple and controllable music generation’, in \textit{Advances in Neural Information Processing Systems}, 2023.

\bibitem{b29} D. von Rütte, L. Biggio, Y. Kilcher, and T. Hofmann, ‘Controllable music generation using learned and expert features’, in \textit{International Conference on Learning Representations}, 2023.

\bibitem{b15} A. Roberts, J. Engel, C. Raffel, C. Hawthorne, and D. Eck, ‘A hierarchical latent vector model for learning long-term structure in music’, in \textit{Proceedings of the 35th International Conference on Machine Learning}, 2018.

\bibitem{b16} L. Min, J. Jiang, G. Xia, and J. Zhao, ‘Polyffusion: A diffusion model for polyphonic score generation with internal and external controls’, in ISMIR, 2023.

\bibitem{b12} J. Zhang, G. Fazekas, and C. Saitis, ‘Composer style-specific symbolic music generation using vector quantized discrete diffusion models’, in \textit{IEEE 34th International Workshop on Machine Learning for Signal Processing}, 2024.

\bibitem{b7} O. Ronneberger, P. Fischer, and T. Brox, ‘U-Net: Convolutional networks for biomedical image segmentation’, in \textit{Medical Image Computing and Computer-Assisted Intervention}, vol. 9351, 2015, pp. 234–241.

\bibitem{b22} 
J. Ho and T. Salimans, ‘Classifier-free diffusion guidance’, in \textit{NeurIPS Workshop on Deep Generative Models and Downstream Applications}, 2021.

\bibitem{b28} K. He, X. Zhang, S. Ren, and J. Sun, ‘Deep residual learning for image recognition’, in \textit{IEEE Conference on Computer Vision and Pattern Recognition}, 2016, pp. 770–778.

\bibitem{b23} D. Hendrycks and K. Gimpel, ‘Gaussian error linear units (GELUs)’, 2016, \textit{arXiv preprint} arXiv:1606.08415.

\bibitem{b26} S. Hwang, D. Han, C. Jung, and M. Jeon, ‘WaveDH: Wavelet sub-bands guided convNet for efficient image dehazing’, 2024, \textit{arXiv preprint} arXiv:2404.01604.

\bibitem{b25} X. Gao et al., ‘Efficient multi-scale network with learnable discrete wavelet transform for blind motion deblurring’, in \textit{IEEE/CVF Conference on Computer Vision and Pattern Recognition}, 2024, pp. 2733–2742. 

\bibitem{b27} L. Chen, X. Chu, X. Zhang, and J. Sun, ‘Simple baselines for image restoration’, in European Conference on Computer Vision, 2022, pp. 17-33.

\bibitem{b24} Z. Wang, D. Wang, Y. Zhang, and G. Xia, ‘Learning interpretable representation for controllable polyphonic music generation’, in \textit{Proceedings of the 21nd International Society for Music Information Retrieval Conference}, 2020.

\bibitem{b18} Z. Wang et al., ‘Pop909: A pop-song dataset for music arrangement generation’, in \textit{Proceedings of the 21nd International Society for Music Information Retrieval Conference}, 2020.

\bibitem{b19} K. Choi, C. Hawthorne, I. Simon, M. Dinculescu, and J. Engel, ‘Encoding musical style with transformer autoencoders’, in International Conference on Machine Learning, 2020, pp. 1899-1908.

\bibitem{b2} H. Phung, Q. Dao, and A. Tran, ‘Wavelet diffusion models are fast and scalable image generators’, in \textit{IEEE Conference on Computer Vision and Pattern Recognition}, 2023, pp. 10199–10208.















\end{thebibliography}
\end{document}